# Tracking and imaging of dynamic objects in scattering media by time-reversed adapted-perturbation (TRAP) optical focusing

Cheng Ma, Xiao Xu, Yan Liu and Lihong V. Wang*

Optical Imaging Laboratory, Department of Biomedical Engineering,

Washington University in St. Louis, St. Louis, Missouri 63130-4899

*Corresponding author: lhwang@biomed.wustl.edu

**The ability to steer light propagation inside scattering media has long been sought-after due to its potential widespread applications. To form optical foci inside scattering media, the only feasible strategy is to guide photons by using either implanted[1] or virtual[2-4] guide stars[5]. However, all of these guide stars must be introduced extrinsically, either invasively or by physical contact, limiting the scope of their application. Here, we focus light inside scattering media by employing intrinsic dynamics as guide stars. By time-reversing the perturbed component of the scattered light adaptively, we concentrate light to the origin of the perturbation, where the permittivity varied spontaneously. We demonstrate dynamic light focusing onto moving targets and imaging of a time-variant object obscured by highly scattering media, without invasiveness and physical contact. Anticipated applications include all-weather optical communication with airplanes or satellites, tracking vehicles in thick fogs, and imaging and photoablation of angiogenic vessels in tumors.**

To focus light deep into scattering media such as clouds, fogs, and biological tissues, conventional manipulation of its phase is no longer feasible due to random scattering. As the propagation distance increases, the number of unscattered photons decays exponentially and becomes negligible beyond one transport mean free path ($l_t'$)[6]; therefore, control of light is limited to a superficial depth of $l_t'$. Obviously, it is highly desirable to break through this optical diffusion limit.

One can tune the scattered light in phase to maximize the spatial[7] or spatiotemporal[8] energy density at a desired location. By means of iterative wavefront shaping[7-12], a large number of spatial modes are controlled to focus through[7,11], or into[1,13,14], a scattering medium at the target location, where a guide star is placed to generate feedback. However, these methods work by sequentially probing the input modes, an approach which remains impractically slow when the controlled degrees of freedom are many. A faster approach is to time-reverse (by phase-conjugating monochromatic light[15]) the scattered photons from a guide star back to their origin. To this end, nonlinear nanoparticles[16] and fluorescent beads[17] have served as implanted guide stars, and focused ultrasound[2-4,18] has functioned as a virtual one. However, physical implantation is invasive and inflexible, and the virtual alternative requires physical contact for acoustic coupling and low loss at a sufficiently high acoustic frequency, which is unfavorable or infeasible for many scattering media, such as wounds, clouds, or fogs.

Here, we explore the feasibility and potential applications of a new technology, named time-reversed adapted-perturbation (TRAP) optical focusing, which offers a strategy for passively guiding light inside scattering media without invasive implantation or physical contact. It works by remotely probing the scattering medium and employing spontaneous motions or intrinsic permittivity variations of the target as the guide stars. Although our analysis centers on absorption, generalization to the full permittivity tensor is straightforward (see Supplementary Discussion). The TRAP process comprises two steps. In step 1, the scattered light field distributions $\mathbf{E}_1$ and $\mathbf{E}_2$, defined at a location exterior to the scattering region, are probed at instants $t_1$ and $t_2$ while the target absorption changes from $\mu(\mathbf{r}, t_1)$ to $\mu(\mathbf{r}, t_2)$, as depicted in Fig. 1 (a) and (b), and the rest of the medium stays unchanged. The process is written mathematically as $\mathbf{E}_1 = \mathbf{T}\mathbf{E}_1^t, \mathbf{E}_2 = \mathbf{T}\mathbf{E}_2^t$, where $\mathbf{E}_i^t$ ($i = 1,2$) denotes the field distribution on the target plane at instant $t_i$, and $\mathbf{T}$ is a transmission matrix characterizing the medium[19,20]. Subtracting the above equations yields the field perturbation $\Delta \mathbf{E} = \mathbf{E}_2 - \mathbf{E}_1 = \mathbf{T}\Delta \mathbf{E}^t$. As $\mathbf{E}_1$ and $\mathbf{E}_2$ are recorded interferometrically at two instants, instability of the two arms of the interferometer can lead to an overall random phase offset. If the phase of $\mathbf{E}_1$ is adapted by minimizing the norm $\|\Delta \mathbf{E}\|$ (see Supplementary Methods), $\Delta \mathbf{E}^t$ is null anywhere except at the target location $\mathbf{r}$. In step 2, as shown in Fig. 1(c), focusing on the target position is subsequently achieved when the phase-adapted perturbation $\Delta \mathbf{E}$ is phase-conjugated: $\mathbf{T}^T \Delta \mathbf{E}^* = \left(\mathbf{T}^\dagger \Delta \mathbf{E}\right)^* \approx \left(\Delta \mathbf{E}^t\right)^*$, where the superscripts "T" and "†" denote transpose and conjugate transpose, respectively, and $\mathbf{T}^\dagger \mathbf{T} \approx \mathbf{I}$ ($\mathbf{I}$ is the identity matrix) is assumed[19]. In the special case of a moving target inside the scattering region, the perturbation satisfies $\delta\mu(\mathbf{r}_1) = -\delta\mu(\mathbf{r}_2) \neq 0$, where $\delta$ denotes the change from $t_1$ to $t_2$. Under such circumstances, phase-conjugating $\Delta \mathbf{E}$ focuses light simultaneously on the old and the new target locations $\mathbf{r}_1$ and $\mathbf{r}_2$ (see Supplementary Movie 1 for details).

To guarantee high energy gain, digital optical phase conjugation (DOPC)[3,4] is employed in TRAP focusing. In the probing mode (Fig. 1 (d)), the scattered electromagnetic field distribution on a phase-only spatial light modulator (SLM) is measured via digital phase-shifting holography (see Methods). The virtual difference field $\Delta \mathbf{E}$ is synthesized by subtracting the fields acquired before and after the perturbation and then optimizing the phase offset. A phase map $\boldsymbol{\Phi}$ is then extracted from $\Delta \mathbf{E}$ and subsequently conjugated to yield the TRAP phase map ($-\boldsymbol{\Phi}$). In the focusing mode (Fig.1 (e)), an intense planar reading beam is transformed into a time-reversed beam after reflecting off the SLM, and converges onto the target (See Methods and Supplementary Fig. S1 for system setup). Speckle-scale focusing between two optical diffusers is demonstrated by perturbing the medium using a target particle, as shown in Fig. 1 (f) (the cross-sectional intensity distributions along *y* and *x* are shown in Fig. 1 (g) and (h), respectively), and a moving particle induces focusing at two points, as shown in Fig. 1 (i) (see Supplementary Methods for details).

An attractive feature of TRAP is the ability to dynamically focus light onto a moving target hidden inside a scattering medium. As the target moves, a remote sensor keeps taking snapshots of the scattered electromagnetic field $\mathbf{E}_i$ (index *i* denotes the current target location) and generating phase maps from $\Delta\mathbf{E}_i^* = \mathbf{E}_i^* - \mathbf{E}_0^*$, where $\mathbf{E}_0$ is obtained when the target is absent. Repetitively reading the newly updated phase map ensures focusing onto the target, given an adequately short time lag between phase conjugation and the latest snapshot. The concept is demonstrated in Fig. 2, where light focuses dynamically from a "ground station" onto an airplane flying inside clouds (Fig. 2(a)). In the experiment shown in Fig. 2(b), an absorptive airplane pattern, printed on a transparency, was continuously translated between two optical diffusers. Meanwhile, the intensity distribution on the target was projected onto a camera by a beamsplitter to monitor the TRAP focus in real time. The light distribution "below the clouds", as measured on the SLM surface, manifested a random speckle pattern (Fig. 2 (d)), indicating complete loss of the wavefront information. In salient contrast, as shown in Fig. 2 (c), light was dynamically focused onto the moving target as if scattering were totally suppressed (see Methods for the experimental details and Supplementary Movie 2 for the dynamic focusing process). This demonstrated feature is especially useful in laser guidance and tracking, and for free-space optical communications in highly scattering media.

In deep tissue imaging, TRAP can enhance the signal and contrast by redistributing and concentrating light on the targets, hence extending the imaging depth. For example, red blood cells (RBCs) are extensively employed as endogenous contrast agents. They provide important physiological information, such as metabolism[21], and their constant motion can be employed to create ideal virtual guide stars. At two instants $t_1$ and $t_2$, RBCs form two sets of spatial patterns within blood vessels. $S_i$ denotes a set of RBC positions at instant $i$ (i = 1,2). If $\Delta t = |t_1 - t_2|$ is selected to be between $\tau_{RBC}$ and $\tau_{tissue}$ (the field decorrelation times associated with the RBC and tissue movements, respectively), TRAP generates optical foci at both $S_1$ and $S_2$, and thus concentrates energy on the RBCs. The concept was demonstrated by the set-up shown in Fig. 3 (a), where a blood-filled translucent silicone tube (300 μm inner diameter) was buried between two pieces of *ex vivo* chicken tissue samples, each 2.5 mm thick (~ 2.5 $l_t'$ [4], see Supplementary Methods for the sample preparation). During the TRAP procedure, $S_1$ and $S_2$ were realized by flowing diluted bovine blood (see Supplementary Methods for the sample preparation) in the tube. The comparative focal light intensity distributions shown in Fig. 3 (b) and (c) demonstrate that TRAP achieved significant energy enhancement at the target region. The enhancement factor, according to Fig. 3(d), was approximately 300%, limited by experimental constraints (see Supplementary Methods for details). Furthermore, when the vessels formed a three-dimensional (3D) pattern, TRAP faithfully shaped light onto the corresponding structure, as shown in Fig. 3 (e) to (g) (see Methods and Supplementary Movie 3 for a 3D visualization).

Compared to schemes employing ultrasonic tagging[2-4], the major advantage of TRAP is efficiency. Ultrasound fundamentally lacks specificity. For targeted light delivery (such as in photodynamic therapy), an image has to be acquired before the region of interest is determined. Moreover, the ultrasonic tagging approach concentrates light only at the ultrasonic beam, whereas TRAP focusing simultaneously enhances energy deposition onto the perturbed target in the entire field of view. Such a unique capability can be extremely useful in many applications such as phototherapy of port wine stains and photoacoustic computed tomography of blood vessels[21].

A time-varying object can also be optically imaged by TRAP. In the experiment shown in Fig. 4 (a), an absorptive target (the number "4") hidden between two optical diffusers was completely invisible externally. When the target moved into the scattering region being probed, its TRAP pattern copied and spatially overlapped itself. We employed the memory effect[22] to scan the TRAP pattern across the target laterally (in *x* and *y*) by stepping through different phase gradients on the SLM while the transmitted light was monitored by a camera located outside the scattering region. The autocorrelation of the object, shown in Fig. 4 (b), was assessed after mathematical manipulations (see Supplementary Methods for details), and the shape of the object was readily obtained by an iterative algorithm[23], as depicted in Fig. 4 (c). In comparison to previous approaches based on the memory effect for imaging through or inside scattering media[16,24,25], TRAP focusing is completely non-invasive and label-free, and it images only the perturbed regions. This scheme will find its niche in vision in dynamic foggy or smoky environments, and imaging of angiogenic vessels in tumor.

Both the speed and energy enhancement ratio of the current system can be improved. Once faster SLMs and cameras are available, the speed can be improved by several orders of magnitude. Alternatively, an off-axis amplitude holographic configuration is able to capture fast perturbations, as short as 10 ns (see Supplementary Discussion). The energy enhancement ratio at the target locations can be dramatically increased if more spatial modes are phase conjugated (enabled by large pixel count SLMs and cameras[26]). We performed light focusing and imaging between two scattering media, where open access to the "inside" of the combined scattering medium enables convenient validation. This arrangement is, in principle, equivalent to focusing into a scattering medium[18].

We emphasize that absorption is not the only perturbation mechanism—any perturbation to the complex permittivity, such as changes in the refractive index, would work equally well (see Supplementary Discussion). Moreover, TRAP does not necessarily rely solely on endogenous contrast agents. Its capability could be further extended by introducing exogenous agents with controlled motions or absorptions, such as magnetomotive particles[27], voltage-sensitive dyes[28] and photo-switchable dyes and proteins[29,30]. By incorporating such labeling strategies, TRAP focusing could be made even more versatile and powerful.

In summary, the TRAP technology is envisioned to have profound impacts in a wide range of applications where the scattering effect needs to be suppressed, including vision and communication; laser guidance, tracking, and trapping; photoacoustic tomography;

optogenetics; photothermal therapy, and photodynamic therapy. Other wave-related fields can also potentially benefit from the same concept.

**Methods**

**Experimental setup.** The experimental setup used to generate and acquire the experimental data is shown in Supplementary Fig. S1. The source used was a Q-switched frequency-doubled Nd:YAG laser (Elforlight Inc., UK) centered at 532 nm with a pulse duration of 10 ns and a coherence length of 7 mm. The repetition rate of the laser was tunable between 50 Hz and 200 Hz, and the full pulse energy was 0.6 mJ. Before entering the Mach-Zehnder interferometer, the light beam was collimated to a diameter of 2 mm by a beam expander. The power injected into the system was adjustable via a half-wave plate (HWP) paired with a polarizing beamsplitter (PBS), and further attenuated through a neutral density filter. Light was split into a sample beam and a reference beam by another PBS: the ratio between the two beams was controlled by a HWP. The polarization of the reference beam was rotated by $90^\circ$ before passing through an acousto-optic modulator (AOM) which up-shifted its frequency by $\delta f$. The reference beam was further expanded by an afocal lens pair to a diameter of 25 mm and spatially filtered by a pinhole. The sample beam entered a scattering cell composed of two optical diffusers (DG10-120, Thorlabs, USA) or two pieces of *ex vivo* chicken breast tissue. The reference beam was combined with the scattered light from the sample by a 50:50 beamsplitter (BS1). The path lengths of the reference and sample beams were balanced for strongest interference using two tunable delay lines. The combined beams then entered the DOPC module. Inside the DOPC, the beams were retro-reflected by the surface of a phase-only SLM (vis-PLUTO, Holoeye, Germany). The retro-reflected beams were further reflected by a 90:10 (90% transmission) beamsplitter (BS2) to form a speckle pattern on an sCMOS camera (pco. Edge, PCO AG, Germany). The pixels of the camera and the SLM were digitally aligned with ≤1 (pixel) alignment error for optimal phase conjugation quality. Once the scattered fields were measured at two instants, the TRAP phase map was calculated and displayed on the SLM. In the hologram reading mode, the sample beam was blocked. The reference beam passed through BS1 and BS2 and impinged on the SLM. The reflection through the SLM yielded the phase-conjugated beam, which was delivered to the scattering medium after passing through BS2 and BS1. The light intensity distribution on the focal plane inside the two scattering media was monitored either by replacing the target with a CMOS camera (Firefly MV, Point Grey Research, Canada), or by using a BS to reflect the focal pattern onto the camera.

**TRAP phase map measurement.** The frequency difference between the sample and reference beams was set by the AOM modulation frequency, which was $\delta f = 50.0000125$ MHz. The AOM was triggered by a function generator (DG4000, RIGOL Technologies, China) with a sinusoidal waveform at $\delta f$. A delay generator (DG645, Stanford Research Systems, USA), sharing a global time-base with the function generator, produced two channels of pulse sequences at $f_s = 50$ Hz with a proper relative delay to externally trigger and synchronize the laser and the camera. The interferometric beat $\delta f$, when undersampled at $f_s$ by the camera, cycled through four phases—0, $\pi/2$, $\pi$, $3\pi/2$—corresponding to four interferograms $\mathbf{P}_i$ ($i = 1, 2, 3, 4$). During the measurement, the SLM was blanked by displaying a constant phase map (phase value = 0). The complex amplitude of the field was readily obtained by $\mathbf{E} = (\mathbf{P}_1 - \mathbf{P}_3) + i(\mathbf{P}_2 - \mathbf{P}_4)$. After the two complex amplitudes were acquired, they were subtracted to get $\Delta\mathbf{E}$, and the TRAP phase map was calculated by $\mathbf{\Phi} = -\mathrm{Arg}\left[\mathrm{Im}(\Delta\mathbf{E})/\mathrm{Re}(\Delta\mathbf{E})\right]$.

For applications requiring higher measurement accuracy (e.g., the experiment shown in Fig. 1 (f) to (i), where the perturbation was weak), a different signal acquisition scheme was used. The AOM was driven at $\delta f$ = 50.0000001 MHz, and the delay generator triggered the laser and the camera at 200 Hz and 4 Hz, respectively. The sampling frequency of $f_s$ = 4 Hz ensured a phase difference of $\pi/2$ between neighboring frames. The exposure time of the camera was set to 51 ms, during which 10 laser shots were averaged to reduce the noise caused by laser shot-to-shot intensity fluctuations.

**Visualization of dynamic light focusing onto a moving target.** To record the video in Supplementary Movie 2, the experimental setup shown in Fig. 2(b) was used. A transparency printed with an airplane symbol was translated perpendicularly to the light axis by a motorized stage. Before the target "flew in", the electromagnetic field on the SLM surface was measured and stored (as $\mathbf{E}_0$), which was subtracted from $\mathbf{E}_i$ ($i$ = 1, 2, …) measured sequentially at a time interval of 2 seconds. Phase-shifting holography measurement took 0.1 second, and TRAP phase map computation took 0.5 second. The target moved at 130 μm/second. The phase conjugated light between the scattering media was sampled by a CMOS camera in real time, and the camera acquisition was triggered synchronously with the reading laser pulses.

**Measurement of the light intensity distribution in 3D.** Figure 3 (e) was obtained by replacing the tube samples with a beam profiler (Ophir BeamGage, Newport Corporation, USA) during the hologram reading process. The profiler surface was aligned in the *x-y* plane, and stepped through fifty points in the -z direction at 250 μm intervals. Displayed in sequence, the acquired images form a video visualizing the light distribution in 3D (see Supplementary Movie 3). Stacking the images into a three-dimensional matrix (by using

VolView) and integrating along the tube directions reveals the line-shaped foci on the side and bottom walls in Fig. 3 (e). Contour plots of the projected intensities are shown in Fig. 3 (f) and (g).

14  Tay, J. W., Lai, P., Suzuki, Y. & Wang, L. V. Ultrasonically encoded wavefront shaping for focusing into random media. *Sci. Rep.* **4**, (2014).

15  Yaqoob, Z., Psaltis, D., Feld, M. S. & Yang, C. Optical phase conjugation for turbidity suppression in biological samples. *Nat Photon* **2**, 110-115, (2008).

16  Hsieh, C.-L., Pu, Y., Grange, R., Laporte, G. & Psaltis, D. Imaging through turbid layers by scanning the phase conjugated second harmonic radiation from a nanoparticle. *Opt. Express* **18**, 20723-20731 (2010).

17  Vellekoop, I. M., Cui, M. & Changhuei, Y. Digital optical phase conjugation of fluorescence in turbid tissue. *Applied Physics Letters* **101**, 081108-081108-081104, (2012).

18  Judkewitz, B., Wang, Y. M., Horstmeyer, R., Mathy, A. & Yang, C. Speckle-scale focusing in the diffusive regime with time reversal of variance-encoded light (TROVE). *Nat Photon* **7**, 300-305, (2013).

19  Popoff, S. M. *et al.* Measuring the Transmission Matrix in Optics: An Approach to the Study and Control of Light Propagation in Disordered Media. *Physical Review Letters* **104**, 100601 (2010).

20  Kim, M. *et al.* Maximal energy transport through disordered media with the implementation of transmission eigenchannels. *Nat Photon* **6**, 581-585, (2012).

21  Wang, L. V. & Hu, S. Photoacoustic Tomography: In Vivo Imaging from Organelles to Organs. *Science* **335**, 1458-1462, (2012).

22  Freund, I., Rosenbluh, M. & Feng, S. Memory Effects in Propagation of Optical Waves through Disordered Media. *Physical Review Letters* **61**, 2328-2331 (1988).

23  Fienup, J. R. Reconstruction of an object from the modulus of its Fourier transform. *Optics Letters* **3**, 27-29, (1978).

24  Bertolotti, J. *et al.* Non-invasive imaging through opaque scattering layers. *Nature* **491**, 232-234, (2012).

25  Katz, O., Small, E. & Silberberg, Y. Looking around corners and through thin turbid layers in real time with scattered incoherent light. *Nat Photon* **6**, 549-553, (2012).

26  Brady, D. J. *et al.* Multiscale gigapixel photography. *Nature* **486**, 386-389, (2012).

27  Jin, Y., Jia, C., Huang, S.-W., O'Donnell, M. & Gao, X. Multifunctional nanoparticles as coupled contrast agents. *Nat Commun* **1**, 41, (2010).
15

**Acknowledgements** We thank Fengbo Zhou for assistance on the flow control system, Lidai Wang for discussion on the experimental design, Yong Zhou for assistance on the dye solution preparation, and James Ballard for editing the manuscript. This work was sponsored in part by National Institutes of Health grants DP1 EB016986 (NIH Director's Pioneer Award) and R01 CA186567 (NIH Director's Transformative Research Award).

**Author Contributions** C.M. and L.V.W. initiated the project. C.M. and X.X. implemented the DOPC-based system. C.M., X.X. and Y.L. designed and ran the experiments. C.M. wrote the codes for the experiments and simulation, and processed the experimental results. L.V.W. provided overall supervision. All authors involved in writing the manuscript.

**Author information** Reprints and permissions information is available at www.nature.com/reprints. The authors declare no competing financial interests. Correspondence and requests for material should be addressed to L.V.W. (lhwang@biomed.wustl.edu).

**Figure Captions**

**Figure 1. Principle and schematic of TRAP. a-c,** Focusing to a single absorber. The exterior fields at instant $t_1$ (**a**) and $t_2$ (**b**) are measured to calculate $\Delta\mathbf{E}$. $\Delta\mathbf{E}^*$ is produced to promote TRAP focusing (**c**). **d-e,** Principle of DOPC. The light paths follow the solid arrows. In the probing mode (**d**), the exterior fields are measured interferometrically to obtain the TRAP phase map. In the focusing mode (**e**), the SLM displays the synthesized phase map, and the reading beam is turned into a phase-conjugated beam to form a focus. **f**, Light intensity distribution measured on the focal plane when a single focus is formed. **g-h**, Focal intensity profiles sampled along *y* and *x* directions in **f**. Solid lines are Gaussian fits. **i**, Light intensity distribution showing focusing at two points due to a moving absorber. L, lens; BS, beamsplitter.

**Figure 2. Dynamic light focusing onto a moving target hidden inside a scattering medium. a,** Illustration of the concept. Light from a phase conjugate mirror (PCM) is focused onto the moving targets (e.g., an airplane or a satellite) through the cloud. **b,** Experimental arrangement. **c,** Focal light intensity distribution. **d,** Speckle pattern observed on the SLM surface during the probing process. BS: beamsplitter. Scale bars, 500 μm.

**Figure 3. Focusing light onto flowing targets inside thick biological tissue. a,** Experimental configuration. Light is focused onto the tube through tissue 2. The focal intensity distribution is monitored by a CMOS camera. **b-c,** Light intensity distribution on the focal plane with an incorrect phase map (**b,** phase conjugation is disrupted by shifting the phase map by 3 pixels) and a correct phase map (**c**) displayed on the SLM.

The dashed lines highlight the tube boundaries. **d,** Averaged intensity distributions across the tube obtained by integrating horizontally within the region enclosed by the dashed-dot lines in **c.** The baseline is obtained by applying the same integration procedure to **b**. **e,** Visualization of 3D focusing. Two vessels are spatially arranged perpendicularly to each other, separated by $\Delta z = 1$ cm in the axial ($z$) direction. The measured light distribution is projected along the $x$ and $y$ directions to the bottom and right surfaces of the cube, respectively. **f-g**, Intensity contour plots of the front (**f**) and the back (**g**) foci. Scale bar in **c**, 500 μm.

**Figure 4. Imaging of a hidden time-variant object**. **a,** Experimental arrangement. The SLM surface is imaged by a lens onto the surface of turbid medium 1 for memory scanning. The energy transmitted through turbid medium 2 is measured by a CMOS camera. Inset: the target and TRAP patterns. **(i),** Photograph of the target (plotted in absorption to reverse the contrast). **(ii),** Intensity distribution of the TRAP pattern. **b,** The autocorrelation coefficient measured as a function of the incident angle. **c,** The reconstructed object plotted in normalized absorption coefficient. Scale bars, 100 μm.

**Figure 1**

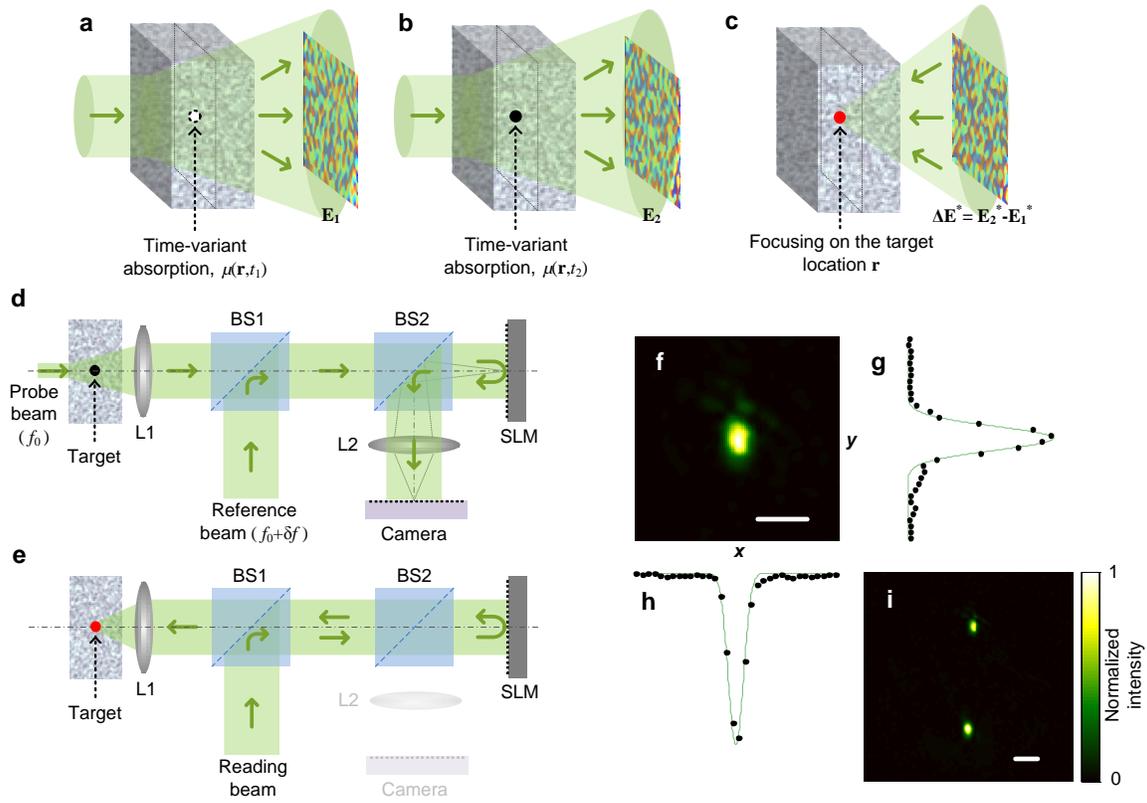

**Figure 1. Principle and schematic of TRAP. a-c,** Focusing to a single absorber. The exterior fields at instant $t_1$ (**a**) and $t_2$ (**b**) are measured to calculate $\Delta\mathbf{E}$. $\Delta\mathbf{E}^*$ is produced to promote TRAP focusing (**c**). **d-e,** Principle of DOPC. The light paths follow the solid arrows. In the probing mode (**d**), the exterior fields are measured interferometrically to obtain the TRAP phase map. In the focusing mode (**e**), the SLM displays the synthesized phase map, and the reading beam is turned into a phase-conjugated beam to form a focus. **f**, Light intensity distribution measured on the focal plane when a single focus is formed. **g-h**, Focal intensity profiles sampled along *y* and *x* directions in **f**. Solid lines are Gaussian fits. **i**, Light intensity distribution showing focusing at two points due to a moving absorber. L, lens; BS, beamsplitter.

**Figure 2**

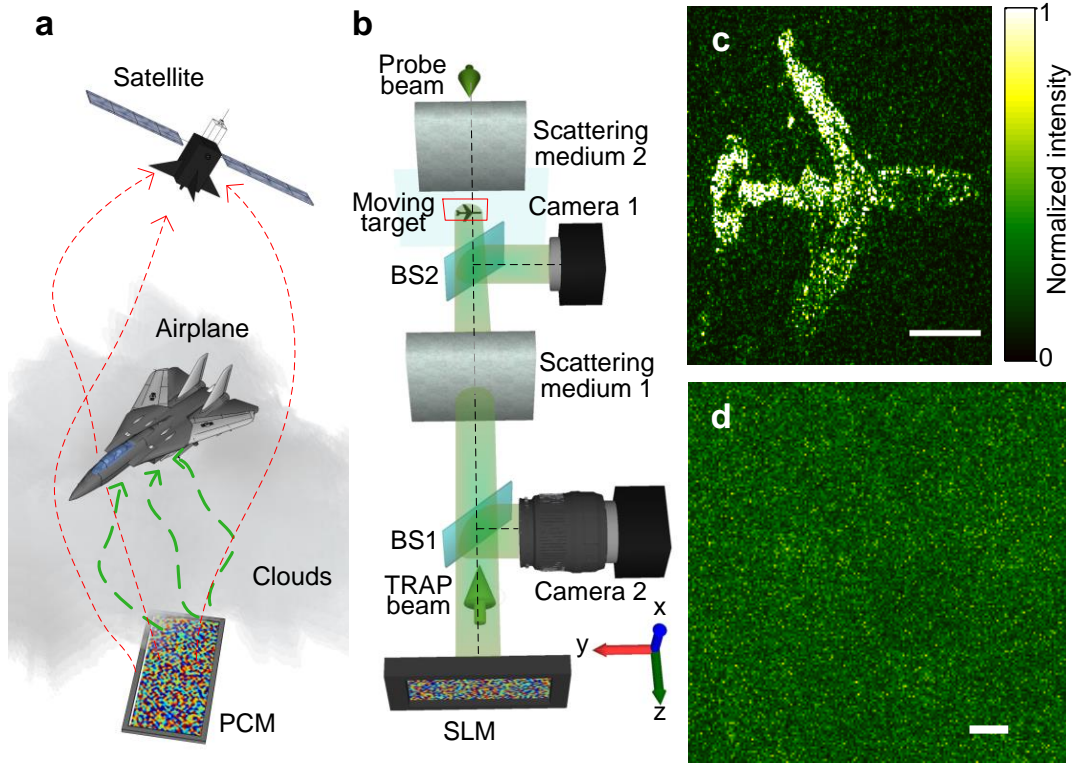

**Figure 2. Dynamic light focusing onto a moving target hidden inside a scattering medium. a,** Illustration of the concept. Light from a phase conjugate mirror (PCM) is focused onto the moving targets (e.g., an airplane or a satellite) through the cloud. **b,** Experimental arrangement. **c,** Focal light intensity distribution. **d,** Speckle pattern observed on the SLM surface during the probing process. BS: beamsplitter. Scale bars, 500 μm.

**Figure 3**

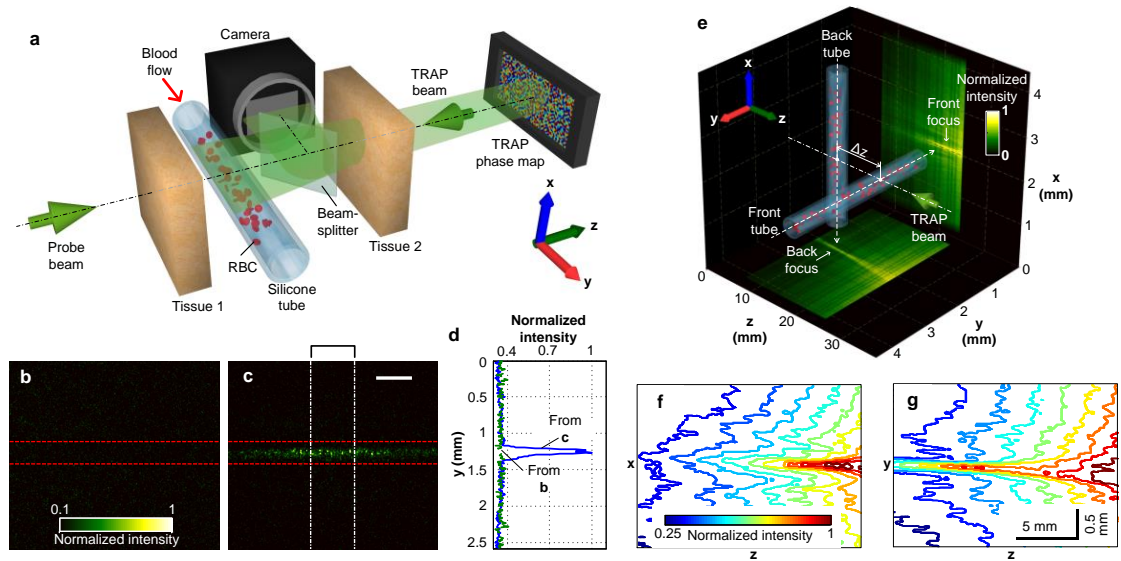

**Figure 3. Focusing light onto flowing targets inside thick biological tissue. a,** Experimental configuration. Light is focused onto the tube through tissue 2. The focal intensity distribution is monitored by a CMOS camera. **b-c,** Light intensity distribution on the focal plane with an incorrect phase map (**b,** phase conjugation is disrupted by shifting the phase map by 3 pixels) and a correct phase map (**c**) displayed on the SLM. The dashed lines highlight the tube boundaries. **d,** Averaged intensity distributions across the tube obtained by integrating horizontally within the region enclosed by the dashed-dot lines in **c.** The baseline is obtained by applying the same integration procedure to **b**. **e,** Visualization of 3D focusing. Two vessels are spatially arranged perpendicularly to each other, separated by Δz = 1 cm in the axial (*z*) direction. The measured light distribution is projected along the *x* and *y* directions to the bottom and right surfaces of the cube, respectively. **f-g**, Intensity contour plots of the front (**f**) and the back (**g**) foci. Scale bar in **c**, 500 μm.

**Figure 4**

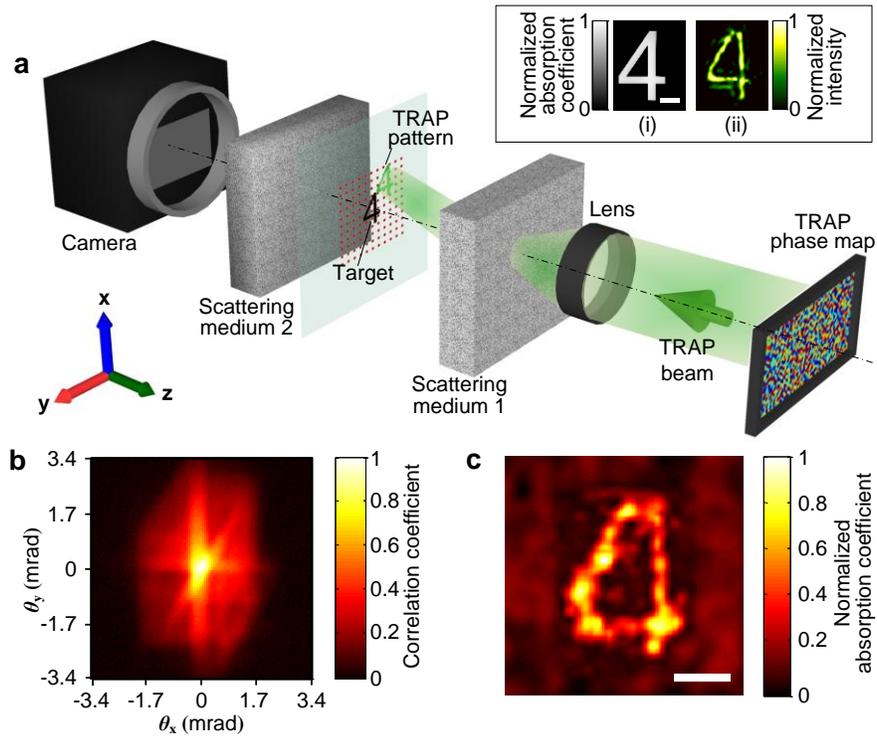

**Figure 4. Imaging of a hidden time-variant object**. **a,** Experimental arrangement. The SLM surface is imaged by a lens onto the surface of turbid medium 1 for memory scanning. The energy transmitted through turbid medium 2 is measured by a CMOS camera. Inset: the target and TRAP patterns. **(i),** Photograph of the target (plotted in absorption to reverse the contrast). **(ii),** Intensity distribution of the TRAP pattern. **b,** The autocorrelation coefficient measured as a function of the incident angle. **c,** The reconstructed object plotted in normalized absorption coefficient. Scale bars, 100 μm.